\documentclass[aoas,nameyear,seceqn,dvips]{arximspdf}
\usepackage{graphics}
\doi{10.1214/09-AOAS307}
\volume{4}
\issue{2}
\pubyear{2010}
\firstpage{893}
\lastpage{915}

\makeatletter
\newcommand{\eqref}[1]{(\ref{#1})}
\newproclaim{remark}{Remark}
\makeatother

\begin{document}
\begin{frontmatter}

\title{Density estimation for grouped data with application to line
transect sampling}
\runtitle{Density estimation for grouped data}

\begin{aug}
\author[a]{\fnms{Woncheol} \snm{Jang}\ead[label=e1]{jang@uga.edu}\corref{}}
\and
\author[b]{\fnms{Ji Meng} \snm{Loh}\ead[label=e2]{meng@stat.columbia.edu}}
\runauthor{W. Jang and J. M. Loh}
\affiliation{University of Georgia and Columbia University}
\address[a]{Department of Epidemiology and Biostatistics\\
University of Georgia\\
Athens, Georgia 30602\\
USA\\
\printead{e1}}
\address[b]{Department of Statistics\\
Columbia University\\
New York, New York 10027\\
USA\\
\printead{e2}}
\end{aug}

% HISTORY:
\received{\smonth{3} \syear{2009}}
\revised{\smonth{10} \syear{2009}}

% ABSTRACT
%
\begin{abstract}
Line transect sampling is a method used to estimate wildlife
populations, with the resulting data often grouped in
intervals. Estimating the density from grouped data can be challenging.
In this paper we propose a kernel density estimator of wildlife
population density for such grouped data. Our
method uses a combined cross-validation and smoothed bootstrap
approach to select the optimal bandwidth for grouped data. Our
simulation study shows that
with the smoothing parameter selected with this method, the estimated
density from
grouped data matches the true density more closely than with other
approaches. Using smoothed bootstrap, we
also construct bias-adjusted confidence intervals
for the value of the density at the boundary. We apply the proposed
method to two grouped data sets, one from a wooden stake study where
the true density is known, and the other from a survey of kangaroos in
Australia.
\end{abstract}

% KEYWORDS
%

\end{frontmatter}
%

%s1 ###
\section{Introduction}
In ecology it is often of great interest to study the abundance
of wildlife populations.
A common approach for estimating the abundance of a biological
population is distance sampling [\citet{barabesi00}; \citet{barabesi02}; \citet{chen96}],
of which line transect sampling is an example.
A~comprehensive review of
distance sampling can be found in \citet{burnham80} and
\citet{buckland01}.

In such studies the detectability of individual
data points often varies with the distance and selection biases are
common. In the basic line transect scheme, for example, a number of
lines of total
length $L$ are randomly placed in the region of interest. Observers
then move along these lines and record the perpendicular distance of
each detected animal from the line. Animals further away from
the lines are more likely to be missed and this can be modeled via a
detection probability function $p(x)$ that represents the conditional
probability of detecting an animal, given that the animal is at a perpendicular
distance $x$ from the line. \citet{buckland01} showed that the density function
of observed distances, denoted $f(x)$, can be obtained from $p(x)$ by
rescaling $p(x)$ to integrate to 1.

In line transect sampling, it is assumed that the line transects
are placed independently of the animal population so that the
animals are distributed uniformly in distance from the lines. The
decrease in observations with distance is then attributed to the
detection function $p(x)$.

Several assumptions about $p(x)$ are also often made.
Since animals are more likely to be missed with increasing distance
from the observer, $p(x)$ is assumed to be montonically decreasing
with $x$. Furthermore, it is assumed that
$p(0)=1$ and $p'(0)=0$, where $p'$
is the derivative of $p$ with respect to $x$, the former representing
the assumption that an animal on the line will not be missed.
By adding the assumption that $\int_0^\infty
p(x)\, dx < \infty$, \citet{burnham76} showed that the average number of
animals per unit
area, $D$, can be estimated with
%
%
%e1.1 ###
\begin{equation}
\hat{D}  = \frac{n\hat{f}(0)}{2L}, \label{eqn:Dhat}
\end{equation}
where $n$ is the number of observations, $L$ is the total length of the
line transects and $\hat{f}(0)$ is an
estimate of $f(0)$.

The rather unintuitive formula (\ref{eqn:Dhat})
can be better understood as follows; suppose that
a strip of width $2w$ and total length $L$ is
surveyed and $n$ animals are detected. The animal density is then given by
\[
D = \frac{n}{2w L P_a},
\]
where $P_a$ is the unconditional detection probability of an
animal in the strip of area~$2wL$, which can be expressed as
\[
P_a=\frac{1}{w}\int^w_0 p(x)\, dx.
\]
With $f(x)= {p(x)}/{\int^w_0 p(x)\, dx},$ and $p(0) =1$, one can show
\[
P_a = \frac{1}{w f(0)},
\]
giving (\ref{eqn:Dhat}).

Due to the difficulty in measuring distances, the
observations are often grouped into convenient distance markers, such
as multiples of five or ten. Thus, estimation of animal populations
using line transect sampling involves estimating a density function
$f$ from grouped data. In particular, the value of the density at the
boundary, specifically, at $x=0$, is of interest.

Various estimation techniques have been proposed for use with line
transect data. \citet{buckland01} introduced parametric modeling of
$f$, of which Fourier series estimators [\citet{burnham80}] form a
subclass. Other methods include kernel density estimation [\citet{chen96}; \citet{mack98}]
and semiparametric methods [\citet{barabesi00}; \citet{barabesi02}]. The reader is asked to refer to the cited works
for details on these methods.

Parametric methods work well if the model is correct. Also, in smaller data
sets, the data may be grouped into as few as 3 or 4 groups. In these cases,
parametric models using covariate information will be useful [\citet{marques03}]. Here, we
focus on nonparametric methods, in particular, on kernel density
estimation using grouped data. In the context of line transect
sampling, the aim will be the estimation of $f(0)$. However, our
proposed method for bandwidth selection in density estimation from
grouped data has applications beyond line transect sampling (see
Section \ref{sect:conclusion}).

Nonparametric estimation of
$f(0)$ has a number of challenges in the grouped data setting:

\begin{enumerate}
\item[1.] \textit{Density estimation from grouped data}: When data are
grouped, using risk function estimators such as the
cross-validation score function to choose the optimal smoothing
parameter can be problematic since the risk function
estimators tend to be monotone
decreasing functions of the smoothing
parameter. As a result, using cross-validation for optimal smoothing
parameter selection may lead to undersmoothing.
\item[2.] \textit{Density estimation at the boundary}: Since
distances are nonnegative, the
support of the density should not include any negative values. To
satisfy this condition, one must modify the original kernel density
estimator to remove any boundary bias.
\item[3.] \textit{Obtaining confidence intervals}: The
standardized form of the nonparametric estimator $\hat f$ can be
expressed as the sum of two terms:
\[
\frac{\hat f(x) - f(x)}{\sqrt{\operatorname{Var}(\hat f(x))}} =
\frac{\hat f(x)
- \mathsf{E}(\hat f(x))}{\sqrt{\operatorname{Var}(\hat f(x))}} +
\frac{\mathsf{E}(\hat f(x)) - f(x)}{\sqrt{\operatorname{Var}(\hat
f(x))}}.
\]
While the first term converges to the standard normal distribution by
the central limit theorem, in nonparametric inference the
second term is not negligible because of the bias-variance
trade-off. Common smoothing techniques require the bias and
the standard error to be of the same order. Therefore, confidence
intervals based on the traditional
form of $\hat f(x) \pm z_{\alpha/2} \sqrt{\operatorname{Var}(\hat
f(x))}$ do not
necessarily achieve the nominal level.
\end{enumerate}

Note that the second and third points above are also common issues in
nonparametric inference for ungrouped data as well. \citet{chen96}
and \citet{mack98} used kernel methods to address these two issues
for ungrouped data. \citet{barabesi02} developed a semiparametric
method for grouped data, but used the traditional form of the
confidence interval for $f(0)$. Optimal bandwidth selection
plays an important
role in addressing the second and third issues whether data are grouped or
not. In this work, we develop an inference procedure that addresses
all three issues together.

Specifically, we propose a combined cross-validation and
smoothed bootstrap procedure to select the
optimal bandwidth in kernel density estimation
with grouped data and to
construct bias-adjusted confidence intervals for the density at the
boundary. To
adjust for the boundary bias, we employ a symmetrization technique
introduced by
\citet{buckland92}. Our methods can be easily extended to
multivariate
cases. We are not aware of any other work that addresses all
aforementioned issues together.

The paper is organized as follows. Section \ref{sect:inference} provides
a brief overview of kernel density estimators and includes a
description of the symmetrization
technique for kernel density estimates at the boundary. In Section \ref
{sect:grouped} we introduce a smoothed bootstrap approach
for bandwidth selection for grouped data. Section \ref{sect:interval} explains
our approach for constructing bias-adjusted confidence intervals for
$f(0)$ and the animal population density $D$.
We present two case studies in
Section \ref{sect:data}, one using data from a wood stake study and the
other from a survey of kangaroos in Australia.
Section
\ref{sect:simstudy} shows the performance of the proposed method in
simulation studies with data generated from artificially constructed
densities commonly used to test kernel density estimators as well as
with a simulated line transect data set.
Concluding remarks follow
in Section \ref{sect:conclusion}.

%s2 ###
\section{Inference for $f(0)$} \label{sect:inference}

Suppose that we have a sample $X_1, \ldots, X_n$ from the density
function $f(x)$. The nonparametric kernel density estimator of $f(x)$
is given by
\[
\hat f_h(x) = \frac{1}{nh}\sum_{i=1}^n K \biggl(\frac{x-X_j}{h}
\biggr),
\]
where $h$ is the bandwidth and $K$ is a bounded, symmetric
kernel function integrating to one. In kernel methods, the choice of
the bandwidth is more crucial than the choice of kernel. The bandwidth
specifies the amount of smoothing applied to the data and controls the
performance of $\hat{f}_h(x)$. For grouped data, the choice of $h$ will
be addressed in Section \ref{sect:grouped}. We use the
Gaussian kernel throughout this paper.

In line transect sampling, the interest is in the value of the density
at the boundary $x=0$, $f(0)$, since this quantity is related to the
animal population density.
It is well known that kernel estimators suffer from high bias near
boundaries [\citet{wasserman05}]. \citet{barabesi00} used local
likelihood density estimation to reduce this boundary bias, and
\citet{barabesi02} extended this approach to grouped data.
We will instead employ the symmetrization technique
used in \citet{chen96}, originally suggested by \citet{buckland92}.
The key idea of the symmetrization technique is to duplicate the data
by reflecting the data about the boundary: we replace each
data value $x_i$ with $x_i$ and its reflection $-x_i$ about 0. Then we
assume that the data consist of values
$y_1, \ldots, y_{2n}$ where $y_{2i-1} = x_i$ and $y_{2i} =-x_i$.
Thus, $X = |Y|$ and we have
\[
f(x) = g(x) +g(-x),
\]
where $g$ is the density of $y$.
The kernel estimator of $g$ is
\[
\hat g_h(x) = \frac{1}{2nh} \sum_{k=1}^{2n}
K\biggl(\frac{x-y_k}{h}\biggr) = \frac{1}{2nh} \sum_{i=1}^n \biggl[
K\biggl(\frac{x-x_i}{h}\biggr) + K\biggl(\frac{x+x_i}{h}
\biggr)\biggr] ,
\]
so that we have, as the kernel estimator of $f(0)$,
\[
\hat f_h(0) = 2 \hat g_h (0) = \frac{1}{nh} \sum_{k=1}^{2n}
K\biggl(\frac{0-y_k}{h}\biggr) = \frac{2}{nh} \sum_{i=1}^{n}
K\biggl(\frac{x_i}{h}\biggr) ,
\]
where the last equality is due to the symmetry of the Gaussian kernel
about zero.

%s3 ###
\section{Bandwidth selector for grouped data} \label{sect:grouped}

Here, we first describe the cross-validation method and the smoothed bootstrap
method for bandwidth selection with ungrouped data. After highlighting
difficulties with using these methods with grouped data, we introduce
a combined cross-validation and smoothed bootstrap strategy that can
deal with such grouped data.

In density estimation, the performance of the density estimate $\hat
f_h$ is
highly sensitive to the choice of the smoothing parameter $h$ and one
often selects the optimal smoothing parameter from observed data
using some criterion of performance. A~common criterion is the risk
function, $R(f,\hat f_h)$, defined to be the expectation of a loss
function, $L(f, \hat f_h)$, often chosen to be the
integrated squared error (ISE):
\[
L(f, \hat f_h) = \operatorname{ISE} = \int[\hat{f}_h(x)-f(x)]^2\, dx.
\]
The risk function can be written as a sum of the squared bias term and the
variance term,
\[
R(f,\hat f_h) = \mathsf{E}( L(f, \hat f_h) ) = \int
\operatorname{bias}^2(\hat{f}_h(x))\, dx + \int\operatorname{Var}(\hat
{f}_h(x))\, dx,
\]
hence, the optimal smoothing parameter is chosen to balance the tradeoff
between the bias and the variance.

The integrated squared estimator (ISE) can be written as
%
%
%e3.1 ###
\begin{eqnarray}\label{eqn:ise}
\operatorname{ISE} &=& \int[\hat{f}_h(x)-f(x)]^2\, dx
\nonumber
\\[-8pt]
\\[-8pt]
\nonumber
&=& \int\hat{f}_h ^2(x) \,dx - 2 \int\hat{f}_h (x) f(x) \,dx + \int f^2
(x)\, dx.
\end{eqnarray}

Since the last term on the right-hand side of (\ref{eqn:ise}) is
independent of $h$, minimizing ISE
is equivalent to minimizing the first two terms.
As $f$ is not known, the middle term has to be estimated, usually by
cross-validation
or bootstrap.

With cross-validation, this middle term is estimated by
\[
- \frac{2}{n} \sum_{i=1}^n\hat
f_h^{-i} (X_i),
\]
where
\[
\hat f_h^{-i} (x) = \frac{1}{h(n-1)} \sum_{j \ne i}^n
K\biggl(\frac{x-X_j}{h}\biggr)
\]
is the kernel estimate of $f$ with the $i$th data point removed. Thus,
the cross-validation function is
\[
\operatorname{CV}(h) = \int\hat f_h^2 (x)\, dx - \frac{2}{n} \sum_{i=1}^n\hat
f_h^{-i} (X_i) ,
\]
and the value of $h$ that minimizes this function is chosen as the
bandwidth. An asymptotic justification of the cross-validation
procedure can be
found in \mbox{\citet{stone84}}.

The bootstrap is an alternative to cross-validation for
bandwidth selection [\citet{taylor89}]. However, the nonparametric
bootstrap method of
sampling the data points with replacement
and obtaining bootstrap density estimates from the bootstrap
samples cannot capture the bias since the bootstrap estimates are
unbiased. Thus, the smoothed bootstrap is used instead.
This involves obtaining an initial density estimate, $\hat{f}(x;h_{\mathrm{in}})$,
using a \textit{pilot} bandwidth
value $h_{\mathrm{in}}$ and obtaining smoothed bootstrap samples by drawing
samples from
this initial density estimate. The optimal bandwidth is then
the value of $h$ that minimizes
%
%
%e3.2 ###
\begin{equation}
\operatorname{BMISE} (h)  =  \mathsf{E}_S \int[\hat{f}^S(x;h) - \hat
{f}(x;h_{\mathrm{in}})]^2\, dx,
\label{eqn:smoothedbootstrap}
\end{equation}
where $\hat{f}^S$ is the kernel density estimate for the smoothed bootstrap
sample generated from $\hat{f}(x;h_{\mathrm{in}})$ and $\mathsf{E}_S$
represents the mean
over the smoothed bootstrap sampling distribution. This smoothed bootstrap
approach can often perform better
than cross-validation. See the work done by \citet{faraway90} and
\citet{jones96} for the details. \citet{faraway90} recommended choosing
the pilot estimate with cross-validation.

The cross-validation and smoothed bootstrap methods described
above work well with ungrouped data.
In practice, however, data are often binned or rounded to some
extent. Suppose we have a mesh $\{t_k\}_{k=0}^K$ specifying $K$
intervals. The actual data $(X_1, \ldots, X_n)$ is not recorded as
such, but instead is of the form $(v_1, n_1),\ldots,$ $(v_K, n_K)$ where
$n_k$ is the
count in the bin $B_k =[t_{k-1}, t_k)$ and $v_k$ is typically taken to
be the
midpoint of the bin $B_k$. Often the bin size $\delta_k =
t_k-t_{k-1}$ is constant for all $k$, but
this is not required in our proposed method.

It is well known that using the cross-validation function to
select the smoothing parameter leads to undersmoothing if the
proportion of tied data is
larger than some threshold [\citet{silverman86}]. Indeed, any
reasonable risk function estimate may not work as a criterion
for choosing the optimal smoothing parameter if there is
significant overlapping in
the data.

This can be explained heuristically. Since the risk function can be
written as a sum of a squared bias term and a variance
term, selecting the bandwidth that minimizes the risk function
selects the amount of smoothing that balances the bias and the
variance. When data are grouped, however, there are additional
biases so that the squared bias term dominates the variance term in
the risk function. As a smaller bandwidth reduces the bias, using
the risk function produces undersmoothing.
In density estimation, this means that the selected optimal bandwidth
will be
close to~0.

To address this problem, we propose a new bandwidth
selection procedure for kernel density estimation using a
combined cross-validation and smoothed bootstrap
strategy. To estimate $f(0)$ in line transect sampling, we use
this new method for bandwidth
selection (Steps 2--6 below) together with the symmetrization technique
of Section
\ref{sect:inference} (Step 1).

Suppose we have grouped data
with a number of counts within each bin. The estimation procedure
involves the following steps:

\begin{enumerate}
\item Apply the symmetrization technique to $X_i, i=1, \ldots, n$, to
obtain $Y=(Y_1, \ldots, Y_{2n})$. Note that $X$ is binned, so
many of the $X_i$'s, and thus the $Y_i$'s, overlap. This
symmetrization step is performed to reduce the boundary bias in
the estimation of $f(0)$. The remaining steps form the smoothed
bootstrap procedure for bandwidth selection for grouped data.

\item For each bin $k=1, \ldots, K$, generate noise from the uniform
$[-\delta_k/2,
\delta_k /2]$ distribution and add them
to the data points, so that the data points no
longer overlap. Let $Y^U=(Y_1^U, \ldots, Y_{2n}^U)$ denote the new
data.

\item Use cross-validation to calculate the optimal
bandwidth for $Y^{U}$.
\item Repeat Steps 2 and 3 1000 times and let the average of
the optimal bandwidths be $h_{\mathrm{in}}$. An initial density estimate
$\hat g(y;h_{\mathrm{in}})$
is then
obtained using $h_{\mathrm{in}}$ as the pilot bandwidth.

\item Generate $B$ smoothed bootstrap samples $Y^S=(Y_1^S, \ldots, Y_{2n}^S)
$ from $\hat g(y;h_{\mathrm{in}})$.
\item With the smoothed bootstrap samples, evaluate BMISE as a
function of $h$ and find the value of $h$ that minimizes BMISE($h$),
denoted $h_S$.
\item Compute $\hat f(0) = 2 \cdot\hat g(0; h_S).$
\end{enumerate}

In short, we are using a smoothed bootstrap approach with the pilot
bandwidth~$h_{\mathrm{in}}$ found using cross-validation on grouped data with random
noise added to them. Note that the smoothed bootstrap in Step 5
above produces bootstrap samples $Y^S$ from $\hat g(y;h_{\mathrm{in}})$,
and the optimal bandwidth $h_S$ is chosen based on $Y^S$ not~$Y^U$.
Thus, the choice of $h_S$ is not directly affected by the
dependence created in the symmetrization step.

\begin{remark} Smoothed bootstrap samples can be generated from
$\hat
g(y;h_{\mathrm{in}})$ by rejection sampling, but can be generated more simply as follows:
\begin{enumerate}
\item Use the naive bootstrap to resample $Y_1^*, \ldots, Y_{2n}^*$
from $Y^U_1, \ldots, Y^U_{2n}$.
\item Generate $z_i$ from $K(\cdot)$. In our case, since we are
using the Gaussian kernel, we generate $z_i$ from the standard
normal.
\item Set $Y^S_i = Y_i^* + h_{\mathrm{in}} \cdot z_i$ for $i=1,\ldots, 2n$.
\end{enumerate}
\end{remark}

\begin{remark} A referee suggested using a different noise
distribution than the uniform in Step 2 above, specifically, that the
noise distribution be proportional to the detection function. As our
method is intended for applications besides distance sampling, we
decided not to pursue this here.
\end{remark}

%K(\frac{x-y_k}{h})\\
%& =& \frac{1}{2(n-1)h} \sum_{k \ne i}^n [
% K(\frac{x-x_k}{h}) + K(\frac{x+x_k}{h})] .
%This is done because the symmetrized data is not independent, but each
%pair consisting of a data point and its reflection is independent of
%other pairs.

%s4 ###
\section{Confidence intervals for $f(0)$ and $D$} \label{sect:interval}
In this section we construct bootstrap confidence intervals for $f(0)$
based on the kernel density estimates. Constructing confidence
intervals for densities requires accounting for the bias that is not
captured in the naive bootstrap procedure. \citet{hall92} proposed
two methods to account for the bias: explicit bias estimation and
undersmoothing. The former method involves estimating the
leading term of the bias explicitly to obtain a bias-adjusted bootstrap
$t$-confidence interval. The leading term of the bias is a functional
of the second derivative of $f$ and \citet{hall92} suggested using a
plug-in kernel estimator of the derivative. In the undersmoothing method,
a sub-optimal bandwidth of a smaller order than the optimal
bandwidth is
chosen to make $[\mathsf{E}(\hat f (x)) - f(x)]/\sqrt{\operatorname
{\mathsf{Var}}(\hat f(x))}$
negligible.

Assuming that the maximum number $d$ for which the $d$th derivative,
$f^{(d)}$, exists and is known, \citet{hall92} compared the two approaches
and recommended the undersmoothing method. However, in practice, the
value of $d$ is usually unknown. Furthermore, there are no useful
guidelines for the choice of the plug-in kernel estimator of
the derivative and the amount of
undersmoothing.

Thus, we
propose using smoothed
bootstrap to estimate the bias of the kernel density estimate and
construct several confidence intervals based on
our smoothed bootstrap procedure. These confidence intervals are based
on studentized and nonstudentized pivot statistics. We use these
confidence intervals in our
simulation and case studies.

To construct confidence intervals, we first follow Steps 1 to 7 in
Section \ref{sect:grouped} to generate smoothed resamples.
For smoothed resample $X_b^S, b=1,\ldots, B$, define
\[
\hat{f}_b^S(x;h_S) = \frac{1}{nh_S}\sum_{i=1}^{n}
K\biggl(\frac{x-X_{i,b}^S}{h_S}\biggr) .
\]
Define the pivot statistic
$R_{n,b}^S (x)= \hat f^S_b (x; h_S) - \hat
f(x;h_{\mathrm{in}})$. If we let $r^S_\alpha$ denote the $\alpha$ sample
quantile of $(R_{n,1}^S, \ldots, R_{n,B}^S)$, then a $100(1-\alpha)$\%
bootstrap pivot
confidence interval for $f(x)$ is
%
%
%e4.1 ###
\begin{equation}
\bigl( \hat f(x;h_S) - r^S_{1-\alpha/2}, \hat f(x;h_S) - r^S_{\alpha/2}\bigr).
\label{eqn:conf1}
\end{equation}
\citet{faraway90} used a similar pivot statistic to construct
simultaneous confidence bands for $f$.

An alternative is to construct confidence intervals based on a
studentized version of the above pivot statistic. It is known that
studentized confidence intervals are more accurate since these
intervals are second-order accurate [\citet{wasserman05}].

With a suitable estimator $\hat{\sigma}_b^S$ of the standard deviation
$\sigma(x)$ of $\hat{f}(x)$, we can use the studentized pivot
\[
U_{n,b}^S = \frac{\hat f_b^S (x;h_S) - \hat f(x;h_{\mathrm{in}})}{\hat\sigma
_b^S (x)},
\]
yielding a $100(1-\alpha)$\% bootstrap studentized pivotal interval
%
%
%e4.2 ###
\begin{equation}
\bigl(\hat f(x;h_S) - u^S_{1-\alpha/2} \hat\sigma(x), \hat f(x;h_S) -
u^S_{\alpha/2} \hat\sigma(x)\bigr), \label{eqn:conf3}
\end{equation}
where $u^S_{\alpha}$ is the $\alpha$ sample quantile of $(U_{n,1}^S,
\ldots, U_{n,B}^S)$. Please refer to the \hyperref[sect:appendix]{Appendix} for details on how to
obtain $\hat{\sigma}_b^S$.

To construct a confidence
interval of $D$, we use equation (\ref{eqn:Dhat}). Here there is
additional variability in $\hat{D}$ due to $n$ being random. \citet
{buckland01} showed that the standard error of $\hat D$
is given by
\[
\hat\sigma_{D}= \hat D \sqrt{ \biggl(\frac{\operatorname{Var}(n)}{n^2} +
\biggl[\frac{\hat\sigma(0)}{\hat f(0)}\biggr]^2 \biggr) }.
\]
If we follow the common practice of using the Poisson for the
distribution of $n$, $\operatorname{Var}(n)$ can be estimated by the value of
$n$, and the above
expression simplifies to
\[
\hat\sigma_{D}= \hat D \sqrt{ \biggl(\frac{1}{n} +
\biggl[\frac{\hat\sigma(0)}{\hat f(0)}\biggr]^2 \biggr) }.
\]
We use this latter formula in our analyses and simulation study.

Using the same approach of defining a studentized pivot statistic, we
get the following $100(1-\alpha) \%$ confidence interval for $D$:
\[
( \hat D - w^S_{1-\alpha/2} \hat\sigma_D, \hat D - w^S_{\alpha/2}
\hat
\sigma_D),
\]
where $w^S_{\alpha}$ is the $\alpha$ sample quantile of the pivot
statistics $W_{n,1}^S, \ldots, W_{n,B}^S$ computed from the bootstrap
sample, and $\hat{\sigma}_D$ is as given above. See the \hyperref[sect:appendix]{Appendix} for details.

%s5 ###
\section{Case studies} \label{sect:data}

We next look at two case studies, one involving a wooden stake data set
and the other a survey of kangaroos in Australia. All computation, including
implementation of our smoothed bootstrap method, was done with the R
statistical language [\citet{Rcore08}]. The code will be available as
supplemental material at the \textit{Annals of Applied Statistics} website.

%s5.1 ###
\subsection{Stake data in Utah} \label{subsect:stakedata}

We consider here a wooden stakes data set from Logan, Utah, which was
also analyzed by
\citet{burnham80},  \citet{barabesi00}, \citet{barabesi02}. This
data set was collected as part of a larger study on line transect sampling.
In particular, 150 wooden stakes were put within 20 m of a transect
line in a meadow near Logan, Utah. The length of the transect line was
1000 m and the actual density of stakes was known to be $D=37.5$ stakes$/$hectare.
An observer walked along the transect line and searched visually for
the stakes.

%f1 ###
\begin{figure}[b]

\includegraphics{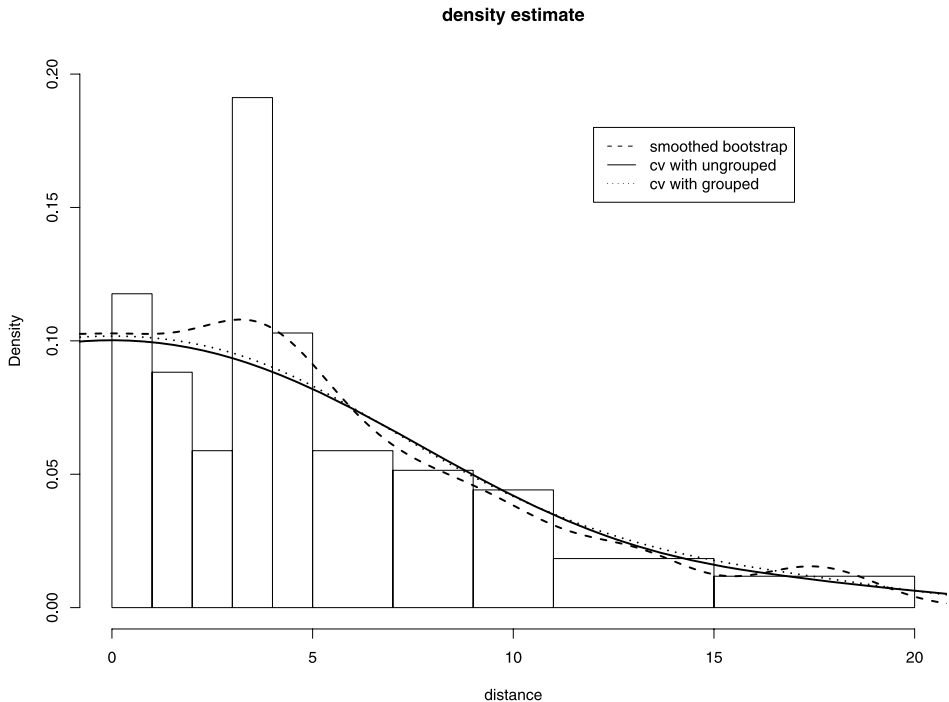}

\caption{Wooden stakes with kernel density estimates.}
\label{fig:density5}
\end{figure}

Out of 150 stakes, 68 were observed. The actual perpendicular
distances of the identified stakes from the transect line are given in
Table 6 of \citet{burnham80}. We notice that more than one stake is
found at some distances. In an actual application these distances are
not known,
but are estimated by the observer. With ten distance categories with
end points $1,2,3,4,5,7,9,11,15,20$, the data then consist of counts
of $8,6,4,13,7,8$, $7,6,5,4$ in the ten distance categories.
Note that the intervals do not have the
same length.

Figure \ref{fig:density5} shows a histogram of the
relative frequencies and kernel density estimates with
bandwidths obtained using different selection methods.

The density estimate with the smoothed bootstrap bandwidth
seems a better fit and also yields the estimate $\hat f(0) =0.1033 $,
or $\hat D
= 35.11$, which is closer to the true density $D=37.5$. For both
grouped and ungrouped data, we received the following warning message:
\textit{minimum occurred at one end of the range} from R and the
lower bounds of the bandwidth range were chosen as optimal
bandwidths with cross-validation. With cross-validation bandwidth
selectors based on ungrouped data and grouped data, we found $\hat D
= 34.07$ and $\hat D = 34.58$ respectively.

\citet{burnham80} fit Fourier series models to the ungrouped and
grouped data to obtain confidence intervals for $D$. As pointed out
in \citet{mack98}, the Fourier series method requires specifying a
horizon, the maximum sighting distance, which is not well defined for
grouped data.

\citet{barabesi00} suggested a local likelihood method to make
inference for $f(0)$, but the method is mainly developed for ungrouped
data. \citet{barabesi02} used density estimation with local least
squares to
obtain estimates for $D$, with bandwidth chosen using a plug-in
method. While their method can be used for grouped data, the resulting
confidence interval for $f(0)$ does not account for the estimation
bias.

%
%
%t1 ###
\begin{table}
\tabcolsep=0pt
\caption{Confidence intervals for $D$}\label{table:ci}
\begin{tabular*}{\textwidth}{@{\extracolsep{\fill}}p{10cm}c@{}}
\hline
\textbf{Method} & \textbf{95\% interval} \\
\hline
Fourier series method with ungrouped data [\citet{burnham80}] &
(32.28, 45.72)\\
Fourier series method with grouped data [\citet{burnham80}] &
(23.95, 40.90)\\
Local likelihood [\citet{barabesi00}] & (27.20, 52.09)\\
Local least squares [\citet{barabesi02}] & (22.13, 49.25)\\
Smoothed bootstrap & (26.65, 45.57)\\\hline
\end{tabular*}
\end{table}

Table \ref{table:ci} shows the confidence interval we constructed from
the wood stake data using the bootstrap method described in Section
\ref
{sect:interval}.
For comparison, we have also included in the
table confidence intervals obtained from the above-mentioned references.
While all confidence intervals cover the true value
$D=37.5$, there are interesting differences.

First note that the first confidence interval is based
on the ungrouped data and, thus, it is the shortest. Information is lost
when data are grouped, and it is expected that the other confidence
intervals will not be as precise. The second confidence interval is
based on the Fourier series method,
applied to the grouped data. This is a parametric method based on the
maximum likelihood estimator and the length of this confidence
interval is shorter than other confidence intervals using the grouped
data. The confidence interval on the third
line is based on a method developed for ungrouped data but applied to the
grouped data. Notice the much wider confidence interval obtained as a
result. While the fourth confidence
interval is valid, it fails to consider the estimation
bias. Note that our confidence interval is shorter than other
nonparametric confidence intervals.

%s5.2 ###
\subsection{Kangaroo survey data from Australia} \label{subsect:kangaroo}

Southwell and Weaver (\citeyear{southwell93}) compared various density estimation
techniques for line transect data using a data set of kangaroo
sightings collected at two locations in Australia, Wallaby Creek and
Tidbinbilla Nature Reserve.

The line-transect work was conducted in a 1.5 km$^2$ region in
Wallaby Creek and in a 0.2 km$^2$ region in
Tidbinbilla Nature Reserve.
At each site, a grid of equally-spaced parallel
lines were marked, 100~m and 50~m apart respectively at Wallaby Creek and
Tidbinbilla Nature Reserve.
An observation session would consist of first
randomly selecting a transect and a direction. An observer would
traverse that transect, then another line transect 400~m (Wallaby
Creek) or 200~m (Tidbinbilla) away, and so on, alternating the
direction with each subsequent line transect. Each observation session
would focus on a particular species,
the eastern grey kangaroo
(\textit{Macropus giganteus}) or red-necked wallaby
(\textit{M. rufogriseus}) in Wallaby Creek and the red kangaroo
(\textit{M. rufus}) in Tidbinbilla.

The kangaroos at both locations were used to the presence of
humans. This allowed the line transects to be more closely spaced than
would normally be done. Furthermore, it is also then relatively
straightforward to perform a census of the kangaroo populations. Thus,
the true kangaroo population sizes are known, and serve as a point of
comparison for the line-transect estimation techniques.
Here, we will only use their data on sightings of the eastern grey
kangaroo in Wallaby Creek.

%f2 ###
\begin{figure}[b]

\includegraphics{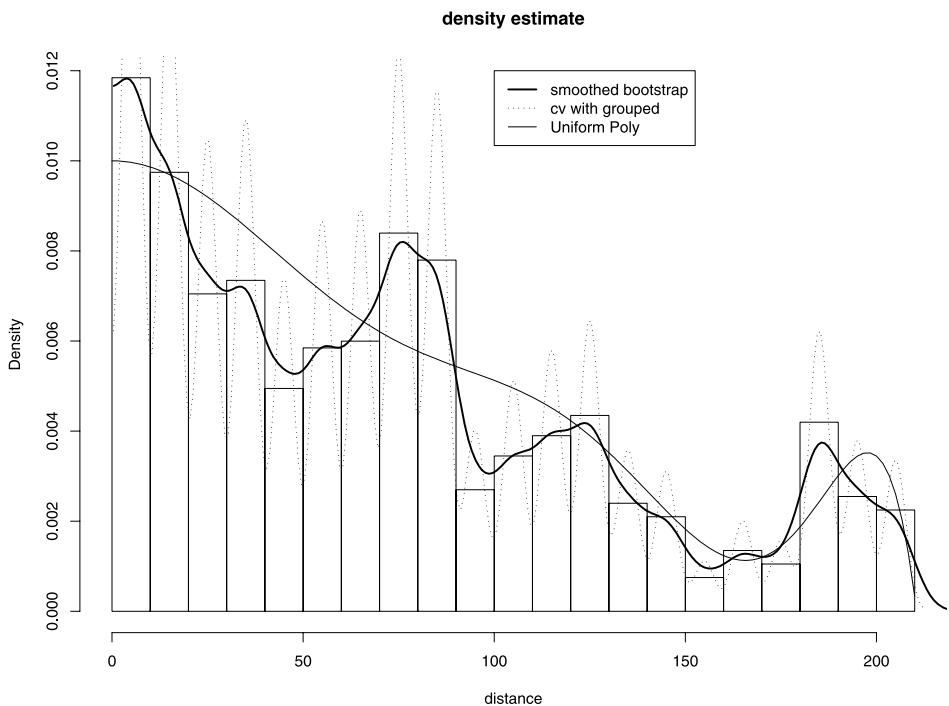}

\caption{Eastern grey kangaroo in Wallaby Creek with kernel density estimates.}
\label{fig:density6}
\end{figure}

Figure \ref{fig:density6} shows a histogram of the eastern grey
kangaroo data, together with kernel density estimates obtained using
optimal bandwidths selected using cross-validation on the grouped data
and using our smoothed bootstrap method. Also shown is a density
estimate obtained using the Distance software program [\citet{thomas09}].
This density estimate was obtained from the model with a
Uniform key function and polynomial adjustment to the tails. This model
was selected from among the other alternatives using AIC as the criterion.

The cross-validation approach yields a density that essentially has a
peak at every bin, while the density obtained with the Distance
software program suggests that too much smoothing may have been
applied. The density estimates obtained from the models with the next
two smallest AIC values, using hazard-rate and half-normal key
functions with cosine adjustments, also suggest over-smoothing (not shown).
The density estimate based on our smoothed bootstrap approach attains a
better fit to the data, with a good balance between smoothing and
retaining the peaks.

The true density D is known to be 44 animals per km$^2$.
Using the smoothed bootstrap approach, we obtained $\hat D =43.71$ and
a 95\% confidence interval of (37.63, 50.51). For comparison, with the
line transect estimate based on the Uniform model with polynomial
adjustment, we have $\hat D = 39.16$ with 95\% confidence interval
(34.91, 43.94).

%s6 ###
\section{Simulation} \label{sect:simstudy}

This section contains two parts.
Section \ref{subsect:simpart2} studies the performance of the bandwidth
selection procedure together with the symmetrization technique for
estimating $f(0)$ using simulated line transect data.

As our method is applicable to areas beyond line transect
sampling, it is of interest to explore its performance under a
variety of settings. In Section
\ref{subsect:simpart1} we apply our bandwidth selection method to data
generated
from artificially constructed densities.
These densities are mixtures of normals and while such densities are
not considered likely in real applications, they are nevertheless
commonly used in the
density estimation community to assess different methods. Here, the
aim is to estimate the whole density function using the selected bandwidth.

%s6.1 ###
\subsection{Simulation study 1} \label{subsect:simpart2}

%f3 ###
\begin{figure}

\includegraphics{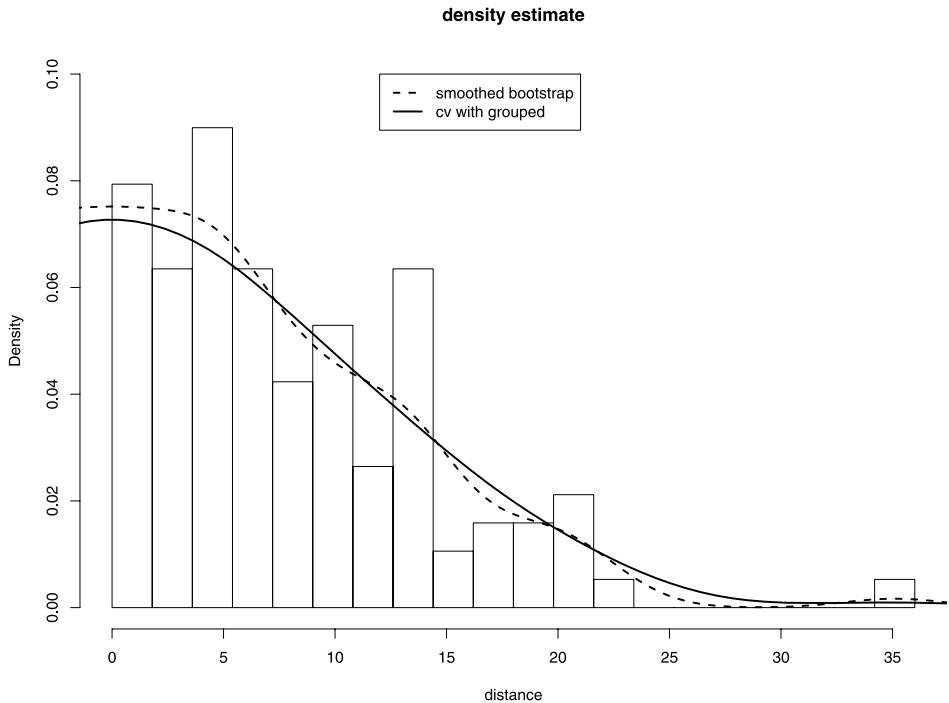}

\caption{Histogram of the simulated transect line sampling data
together with kernel density estimates with bandwidths selected
using smoothed bootstrap and cross-validation with grouped data.}
\label{fig:bucklanddata}\vspace*{-3pt}
\end{figure}

%
%
%t2 ###
\begin{table}[b]
\caption{Estimates of $f(0)$ using smoothed bootstrap,
cross-validation and several parametric\break models fit by
Buckland et al. (\protect\citeyear{buckland01})}\label{table:simpart2}
\begin{tabular*}{250pt}{@{\extracolsep{\fill}}lcc@{}}
\hline
\textbf{Method} & $\bolds{\hat{f}(0)}$ & $\bolds{\hat{D}}$ \\
\hline
Smoothed bootstrap & 0.0751 & 82.14 \\
Cross-validation & 0.0726 & 79.44 \\
Uniform $+$ cosine & 0.0732 & 80.06 \\
Uniform $+$ polynomial & 0.0681 & 74.43 \\
Half-normal $+$ Hermite & 0.0794 & 86.87 \\
Hazard-rate $+$ cosine & 0.0769 & 84.06 \\[3pt]
True value & 0.0798 & 79.79 \\
\hline
\end{tabular*}
\end{table}

Here, we consider a simulated line transect data set that was generated
by \citet{buckland01} for comparing various line transect data
analyses. We briefly describe it below, referring the reader to
\citet{buckland01} for more details.

The data set was simulated so that the assumptions for line transect
sampling hold. It was based on the context of line transect
sampling using 12 parallel line transects of varying lengths within a
region of irregular shape. The
detection function used was the half-normal and the true values of
$f(0)$ and $D$ are 0.0798  m$^{-1}$ (to three significant
figures) and 79.79 objects per km$^2$.
The model was set up so that
the expected number of observations was 96. The simulated data set has
105 observations. The original data set was ungrouped, but was
grouped in various ways by \citet{buckland01} for use with some of the
methods considered there. We use the data which had been grouped into
20 groups of equal width. Figure \ref{fig:bucklanddata} shows a
histogram of the raw data.

We applied smoothed bootstrap and cross-validation for grouped data to
this data set. The resulting density estimates are shown in Figure
\ref{fig:bucklanddata}.
Estimates of $f(0)$ and $D$
are shown in Table~\ref{table:simpart2}. This table also
contains estimates taken from Table~4.2 of \citet{buckland01},
obtained using parametric models fit to the data. These involve
fitting a key
function (uniform, half-normal or hazard-rate) to the data and then
applying an adjustment
(cosine, polynomial or Hermite) to the tails.

Since the true detection function is half-normal, it is not surprising
that the half-normal (with Hermite adjustment) gave an estimate of
$f(0)$ closest to the true value. The estimate obtained with smoothed
bootstrap was closer to the true value than the cross-validation
estimate and the estimates obtained using the uniform model. Note that
to get $D$, the expected value $\mathsf{E}(n)=96$ was used in the formula~(\ref{eqn:Dhat}), while for the estimates, $n=105$ was used.

We note also that the
simulated data had an outlier and \citet{buckland01} recommended
truncating about 5\% of the data, corresponding to dropping
six of the largest observations in this case. After truncation, $\hat
f(0)$ and $\hat D$ were 0.0844 and 87.98 using the half-normal
model with Hermite adjustment. Our method is nonparametric and, hence,
we do not make assumptions about the form of the density. In
particular, our estimate $\hat f(0)$ is robust to outliers in the tails
because the kernel estimate is based on \textit{local} smoothing. Hence,
the presence of outliers does not adversely affect the estimation of
$f(0)$ and our method does not require truncation.

Using the formulas in Section \ref{sect:interval},
we obtained standard errors of 0.017 and 20.36 for $\hat{f}(0)$ and
$\hat{D}$ respectively, assuming the Poisson distribution as the
sampling distribution for $n$. Nominal 95\% confidence intervals for
${f}(0)$ and
$D$, obtained by bootstrap, were $(0.061, 0.092)$ and $(67.18,
101.33)$ respectively.

With a normal approximation approach in \citet{buckland01}, 95\%
confidence intervals for $D$ are (60.14, 113.60) and
(59.36, 116.30) (with truncation).

%s6.2 ###
\subsection{Simulation study 2} \label{subsect:simpart1}

In this section we present results from a simulation study testing the
effectiveness of our bandwidth selection method for estimating the
whole density function from binned data,
a special case of grouped data.

%
%
%t3 ###
\begin{table}[b]
\caption{Parameters for mixture normal densities}\label{table:mixture}
\begin{tabular*}{\textwidth}{@{\extracolsep{\fill}}lcc@{}}
\hline
\textbf{Model} & \textbf{Density} &$\bolds{\sum_{k=1}^K p_k N(\mu_k,\sigma_k^2)}$ \\
\hline
1 & Gaussian & $N(0,1)$ \\[1pt]
2 & Separated bimodal &$\frac{1}{2}N(-\frac{3}{2}, (\frac
{1}{2})^2)+\frac{1}{2}N(\frac{3}{2}, (\frac
{1}{2})^2)$ \\[2pt]
3 & Claw & $\frac{1}{2}N(0,1)+ \sum_{k=0}^4 \frac{1}{10}N
(\frac{k}{2}-1,(\frac{1}{10})^2)$\\[3pt]
4 & Asymmetric claw & $\frac{1}{2}N(0,1)+ \sum_{k=-2}^1
\frac{2^{k-1}}{30}N
(k+\frac{1}{2},(\frac{2^{-k}}{10})^2)$ \\
\hline
\end{tabular*}
\end{table}

%f4 ###
\begin{figure}

\includegraphics{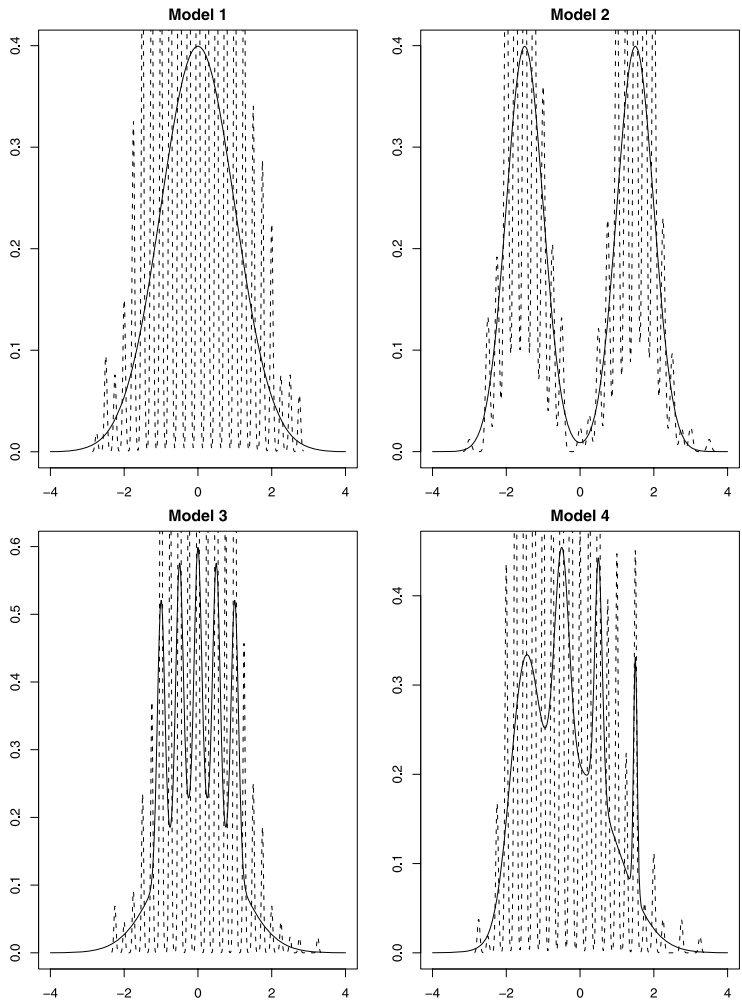}

\caption{Plots showing the true densities of the 4 models we
considered (solid lines) and the kernel density estimates (dashed
lines) using the
cross-validation optimal bandwidths obtained from the binned
data, $h_{\mathrm{cv}}^{\mathrm{bin}}$.}
\label{fig:density1}
\end{figure}

We used four mixture normal densities taken from \citet{wand92}. The
parameters for the mixture densities are shown in
Table \ref{table:mixture} and plots of these densities are shown in Figure
\ref{fig:density1} (solid lines). All simulation studies were
implemented using the R. We generated a
sample of size 500
from each of these densities and binned the data using a bin size of 0.25.

Thus, we have two data sets for each model, one raw and one
binned. Optimal bandwidth selection using cross-validation was
applied to each data set, yielding bandwidth values $h_{\mathrm{cv},i}^{\mathrm{raw}}$
and $h_{\mathrm{cv},i}^{\mathrm{bin}}$ for $i=1, \ldots, 4$, which are
cross-validation optimal bandwidths obtained from the $i$th raw
data set and $i$th binned data set respectively. In R, this is
done using the function \textit{bw.ucv}.
Since this is an optimization problem, a built-in range of
bandwidths is used in the function.
For all models we considered, applying the function to the binned data
sets yielded the
warning message ``\textit{minimum occurred
at one end of the range},'' suggesting that the optimal bandwidths
found using cross-validation are near 0.

%f5 ###
\begin{figure}

\includegraphics{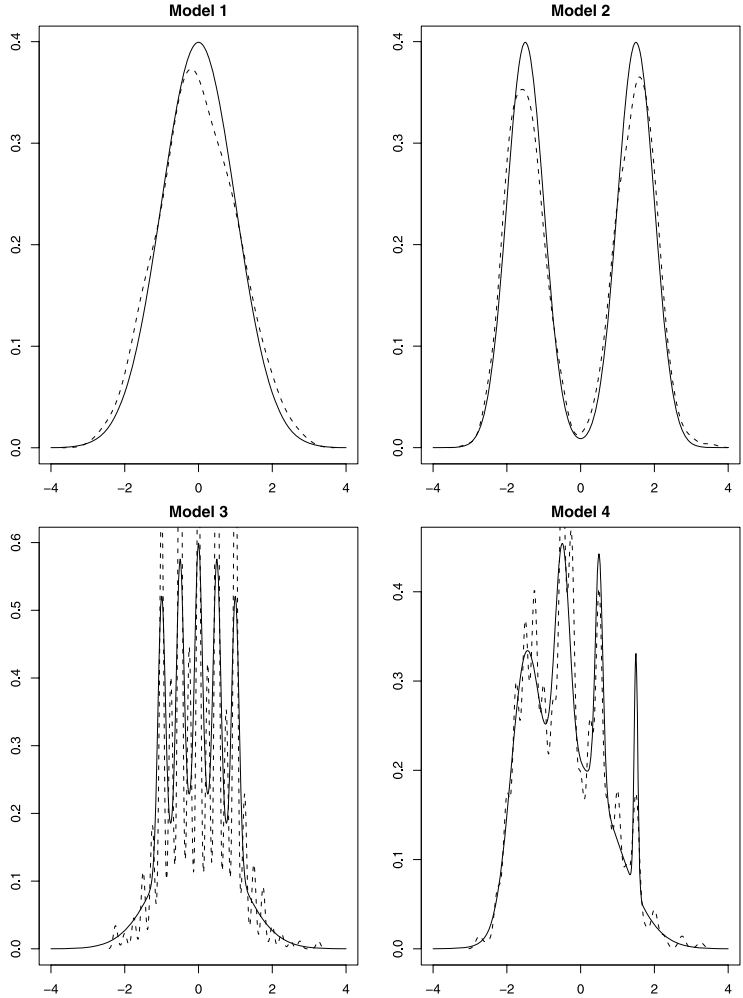}

\caption{Plots showing the true densities of the 4 models we
considered (solid lines) and the kernel density estimates (dashed
lines) using the
cross-validation optimal bandwidths obtained from the original, raw
data, $h_{\mathrm{cv}}^{\mathrm{raw}}$.}
\label{fig:density2}
\end{figure}

%f6 ###
\begin{figure}

\includegraphics{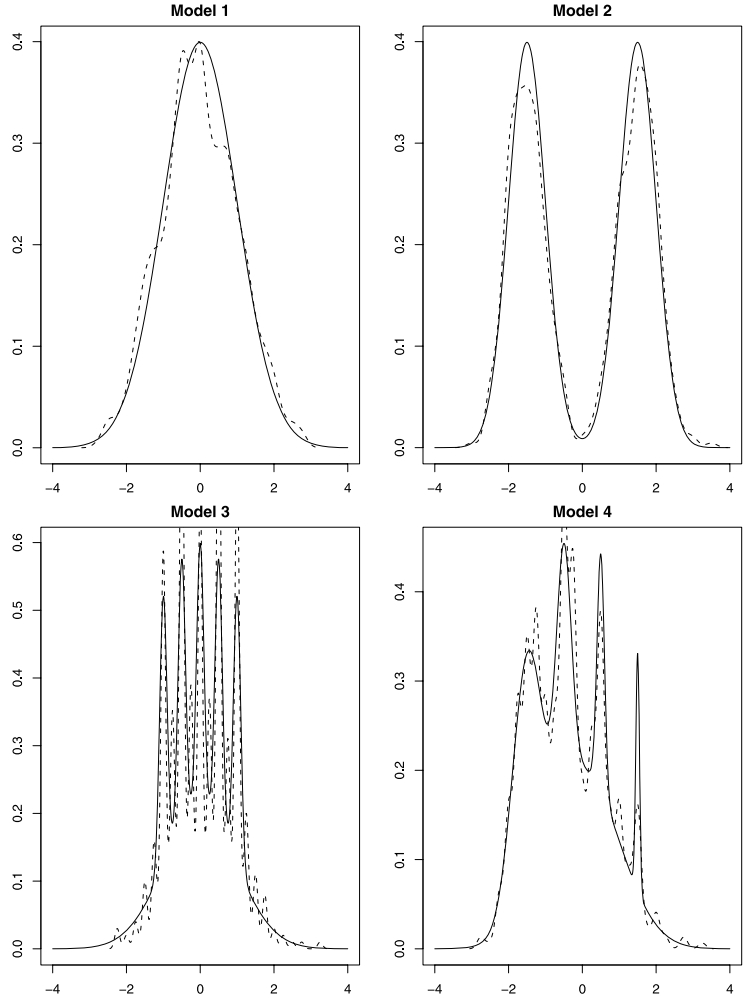}

\caption{Plots showing the true densities of the 4 models we
considered (solid lines) and the kernel density estimates (dashed
lines) using the initial bandwidths $h_{\mathrm{in}}$ obtained from Step \textup{4} of
the smoothed bootstrap procedure.}
\label{fig:density3}
\end{figure}

%f7 ###
\begin{figure}

\includegraphics{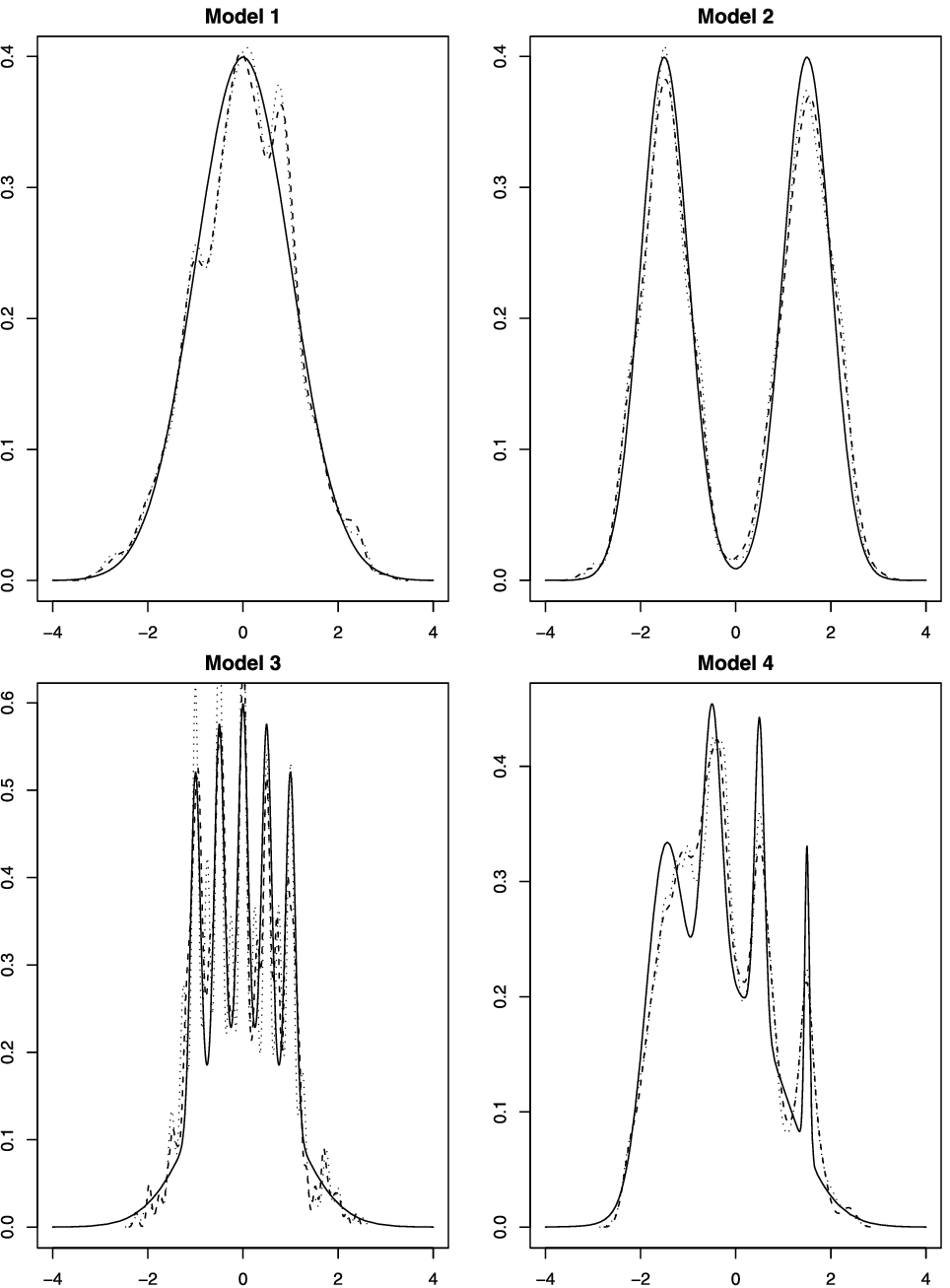}

\caption{Plots showing the true densities of the 4 models we
considered (solid lines) and the kernel density estimates (dashed
lines) using optimal bandwidths selected using smoothed bootstrap,
$h_{S}$. The dashed lines are density estimates using cross-validation
optimal bandwidth from the original, raw data.}
\label{fig:density4}
\end{figure}

Each pair of selected bandwidths are then used with the \textit{binned}
data to obtain kernel density estimates. The results are shown in
Figures \ref{fig:density1} and \ref{fig:density2}, which are
respectively plots of the kernel density estimators using
$h_{\mathrm{cv},i}^{\mathrm{bin}}$ and $h_{\mathrm{cv},i}^{\mathrm{raw}}$.

Figure \ref{fig:density1} shows the problem of using cross-validation
on the binned data to obtain optimal bandwidths. As can be seen,
the selected bandwidths $h_{\mathrm{cv},i}^{\mathrm{bin}}$ are too small,
resulting in severe under-smoothing (dashed lines). In Figure
\ref{fig:density2} we find
that if the underlying true density is relatively smooth (models 1 and
2), using the optimal bandwidths for the raw data,
$h_{\mathrm{cv},i}^{\mathrm{raw}}$, on the
binned data works well. However, if the true density is less
smooth, using $h_{\mathrm{cv},i}^{\mathrm{raw}}$ is not
appropriate for the binned data. Thus, methods such as that
proposed by \citet{chiu91} that aim to obtain approximations to
$h_{\mathrm{cv}}^{\mathrm{raw}}$ may not work if the true density is not
sufficiently smooth.

Figure \ref{fig:density3} shows plots of kernel density estimates
using the pilot bandwidths $h_{\mathrm{in}}$ obtained from Step 4 of our
procedure described in Section \ref{sect:grouped}.
These plots
are similar to those in Figure \ref{fig:density2}, with density
estimates close to the true densities if the true densities are
sufficiently smooth, but with severe under-smoothing otherwise.

Plots of kernel densities estimates using the smoothed bootstrap
optimal bandwidths $h_S$ are shown in Figure
\ref{fig:density4} (dotted lines). For comparison, the kernel density
estimates using $h_{\mathrm{cv}}^{\mathrm{raw}}$ with the raw data (the best
case scenario) are also shown in dashed lines. Note
that the dotted lines are very close to the dashed lines in spite of
some information loss due to the binning. It is
clear that in all the models we considered, the resulting density
estimates are much smoother and closer to the true densities than
using $h_{\mathrm{cv}}^{\mathrm{bin}}$, $h_{\mathrm{cv}}^{\mathrm{raw}}$ or
$h_{\mathrm{in}}$ on
the binned data. Table \ref{table:bandwidth} summarizes the optimal
bandwidth values chosen by different bandwidth selectors.

%
%
%t4 ###
\begin{table}
\caption{Bandwidth comparisons for mixture normal densities}\label{table:bandwidth}
\begin{tabular*}{300pt}{@{\extracolsep{\fill}}lcccc@{}}
\hline
& \textbf{CV with} & \textbf{CV with} & \textbf{Initial} & \textbf{Smoothed} \\
\textbf{Model} & \textbf{raw data} & \textbf{binned data} & \textbf{bandwidth} & \textbf{bootstrap}\\\hline
1 & 0.316 & 0.034 & 0.154 & 0.154\\
2 & 0.191 & 0.054 & 0.148& 0.144\\
3 & 0.058 & 0.028 & 0.066 & 0.074\\
4 & 0.093 & 0.034 & 0.101& 0.112\\
\hline
\end{tabular*}
\end{table}

%s7 ###
\section{Concluding remarks} \label{sect:conclusion}

In this paper we introduced a combined cross-validation and smoothed
bootstrap approach for obtaining kernel density estimates from grouped
data. Our simulation results show that the smoothing parameter found
using our method produced density estimates that matched the true
density most closely compared with competing methods.

In line transect sampling it is the value of the density at the
boundary, specifically $f(0)$, that is of interest, since the estimate
of $f(0)$ is used to estimate the animal population density.
We showed
that the symmetrization technique of \citet{chen96} together with our
bandwidth selection procedure was able to produce good estimates of both
the stake density and the eastern grey kangaroo density.

%In our specific application of line transect sampling, we also tackled
%the problem of obtaining confidence intervals for the density at a
%boundary value. We first adopt the symmetrization technique to account
%for the boundary bias and construct bias-adjusted confidence
%intervals based on smoothing bootstrap. Compared to other confidence
%intervals, our method provide a data driven smoothing parameter and
%accounts for not only the boundary bias, but also the estimation
%bias.

There are some limitations to our method. For application to line
transet sampling, we are restricted to data that is sufficiently large
and grouped into about 10 intervals. With smaller data sizes, the data
may be grouped into as few as 3 or 4 intervals. In such cases, we do
not expect a nonparametric kernel method to work well. Often, a
parametric model involving covariates is used instead.

The methodology developed in this paper has wider potential
application in other scientific areas. For example, economists often
want to make inference for income distributions in developing
counties where only grouped data are available to outsiders
[\citet{perloff07}]. In astronomy, \citet{efron96} applied a
semiparametric density estimator to the estimation for density of
galaxy for which counts on a fine grid are variables. Complex survey
data are another possible application [\citet{bellhouse99}]. We will
explore some of these applications in future work.

% The Appendices part is started with the command \appendix;
% appendix sections are then done as normal sections
\appendix

\begin{appendix}
%s8 ###
\section{Estimation of {$\sigma^2({x})$}}\label{sect:appendix}

We describe how to estimate the variance $\sigma^2(x)$ of $\hat{f}(x)$.
It can be shown that
\begin{eqnarray*}
\sigma^2(x) &=& \frac{1}{nh^2} \int K \biggl(\frac{x-y}{h}\biggr)^2 f(y)
\,dy - \frac{1}{n}\biggl[\frac{1}{h} \int K\biggl(\frac{x-y}{h}
\biggr)f(y)\,dy\biggr],
\end{eqnarray*}
and \citet{hall92} proposed the following estimator of $\sigma(x)$:
\[
[\hat\sigma(x)]^2 = \frac{1}{nh}\biggl[\frac{1}{nh}\sum_{i=1}^n
K
\biggl(\frac{x-X_i}{h}\biggr)^2 - h \hat f(x)^2\biggr].
\]
With our smoothed bootstrap samples, we can estimate the variance by
\[
[\hat{\sigma}^S_b(x)] ^2 = \frac{1}{nh_S} \biggl[
\frac{1}{nh_S} \sum_{i=1}^n K \biggl(\frac{x-X^S_{i,b}}{h_S}
\biggr)^2 -
h_S \hat f_b^S(x;h_S)^2\biggr] ,
\]
and use $\hat{\sigma}_b^S(x)$ in the studentized pivot
statistic $U_{n,b}^S$.

%s9 ###
\section{Confidence interval for $D$}
\citet{chen96} showed that
\[
\frac{\hat D - D - \widehat{\operatorname{bias}}(\hat D)}{\hat\sigma_{D}}
\rightarrow N(0,1),
\]
where $\widehat{\operatorname{bias}}(\hat D) ={n}f^{(2)}(0) h^2/ (2L)$ and
$f^{(2)}$ is the second derivative of $f$.

Based on the same approach that we used to obtain a confidence
interval for $f(0)$, we define a studentized pivot statistic:
\[
W_{n.b}^S = \frac{\hat D_b^S - \hat D_{h_{\mathrm{in}}}}{\hat\sigma^S_{b,D}},
\]
where
\begin{eqnarray*}
\hat D_{h_{\mathrm{in}}} &=&\frac{n \hat f(0; h_{\mathrm{in}})}{2L},\qquad  \hat D^S_b =
\frac{n \hat f^S_b(0; h_S)}{2L},\\
\hat\sigma^S_{b, D}&=& \hat D^S_b \sqrt{
\biggl(\frac{\operatorname{Var}(n)}{n^2} + \biggl[\frac{\hat\sigma_b^2
(0)}{\hat f^S_b(0;h_S)}\biggr]^2 \biggr) }.
\end{eqnarray*}
With $B$ bootstrap samples, we get values $W_{n,1}^S, \ldots,
W_{n,B}^S$. A $100(1-\alpha) \%$ confidence interval for $D$ is then
given by
\[
( \hat D - w^S_{1-\alpha/2} \hat\sigma_D, \hat D - w^S_{\alpha/2}
\hat
\sigma_D),
\]
where $w^S_{\alpha}$ is the $\alpha$ sample quantile of $(W_{n,1}^S,
\ldots, W_{n,B}^S)$.
\end{appendix}

\begin{supplement}[id=suppA]
\stitle{R codes for simulation and case studies}
\slink[doi]{10.1214/09-AOAS307SUPP}
\slink[url]{http://lib.stat.cmu.edu/aoas/307/supplement.zip}
\sdatatype{.zip}
\sdescription{This zip files contains two R scripts for the simulation
and case studies described in \citet{jang09}.}
\end{supplement}

\printaddresses

\end{document}